\documentclass[12pt]{article}

\usepackage{amsfonts,amssymb,amsmath}

\usepackage{graphicx}
\DeclareGraphicsExtensions{.pdf,.png,.jpg}

\begin{document}
\title{\bf Local quantum uncertainty of SU(2) invariant states}
\author{ Esfandyar Faizi $^{a}$
\thanks{E-mail:efaizi@azaruniv.edu}  ,
Bahram Ahansaz $^{a}$
\thanks{E-mail:b.ahansaz@azaruniv.edu}
\\ $^a${\small Physics Department, Azarbaijan shahid madani university, Tabriz, Iran.}} \maketitle

\begin{abstract}
\noindent In this paper, we investigated quantum correlation in $SU(2)$ invariant quantum spin systems by local quantum uncertainty(LQU).
These states are invariant under global rotations of
both subsystems and in real physical systems, such states arise from thermal equilibrium states of isotropic spin systems.
We derive an analytic expression for the LQU of  $(2j + 1)\otimes2$ and $(2j + 1)\otimes3$ quantum spin systems with $SU(2)$ symmetry.
\end{abstract}

\section{Introduction}
The quantum correlation for a quantum state contains entanglement and the other type
of nonclassical correlations\cite{Bennett,Henderson}. Thus, quantum correlations become the subject of intense studies in the last two decades \cite{Modi}. Among the various
researches, it is of great significance to measure the quantum
correlations quantitatively. There are much attention put on
the quantification of bipartite quantum correlations, including quantum discord \cite{Ollivier}, geometric discord \cite{Dakic,Bellomo}, quantum
deficit \cite{Rajagopal}, measurement-induced disturbance \cite{Luo1}, etc, and
each of these measures are useful in particular physical contexts.
Recently, LQU for bipartite quantum systems, as an another measure for quantum correlations, has been proposed by Girolami \cite{Girolami}.
The LQU quantifies the uncertainty in a quantum
state due to measurement of a local observable. Nevertheless, such quantifier has strong reasons to be considered as a faithful measure of quantumness in quantum
states. But due to inherent optimization, finding closed
formula is a difficult problem for most of the correlations measures.
For instance, there is no analytical formulae of quantum discord even for two-qubit quantum systems and
in bipartite systems with higher dimension, the results are known for only some certain
states \cite{Vinjanampathy}.
It is possible to derive closed formula geometric discord and also for quantum discord for Werner and
Isotropic classes of states due to their highly inherent
symmetry in the structures. However, LQU has derived for a large class of arbitrary-dimensional bipartite quantum states \cite{Shuhao}.

More recently, it has been studied that mixed states being invariant under certain joint symmetry operations of the bipartite system.
We consider a slightly more general class of states, rotationally symmetric states, also
known as the $SU(2)$-invariant states. These states are invariant under the global rotations of
both subsystems and their density matrices arise from thermal
equilibrium states of low-dimensional spin systems with
a rotationally invariant Hamiltonian by tracing out all
degrees of freedom except those two spins. $SU(2)$-invariant density matrices of two spins $S_{1}$ and $S_{2}$ are defined to be invariant
under $U_{1} \otimes U_{2}$, and we have $(U_{1} \otimes U_{2})\rho(U_{1}^{\dag} \otimes U_{2}^{\dag})= \rho$ , where $U_{a}=exp(i\vec{\eta}.\vec{S_{a}})$, $a=1,2$ are the usual
rotation operator representation of $SU(2)$ with real parameter $\vec{\eta}$ and $\hbar=1$.
For $SU(2)$ invariant quantum spin systems, negativity is shown to be
necessary and sufficient for separability \cite{Schliemann1,Schliemann2}, and the relative entropy of entanglement
has been analytically calculated \cite{Wang}.
Furthermore, the entanglement of formation (EoF), I-concurrence, I-tangle and convex-roof-extended negativity of the $SU(2)$-invariant states of a spin-$j$ and spin-1/2 \cite{Manne} have been analytically calculated by using the approach in \cite{Vollbrecht}.
Recently, Quantum discord for $SU(2)$-invariant states composed of spin-$j$ and spin-1/2
systems has been analytically calculated in \cite{Cakmak}.
Also, the one-way deficit has been calculated for $SU(2)$-invariant states and show that the one-way deficit is equal to the quantum discord for half-integer $j$ , and is larger
than the quantum discord for integer $j$ \cite{Teng}. Abundance of $SU(2)$ invariant states in real physical systems, make them a good candidate
for utilization in quantum computing protocols.

In the present work the behavior of LQU of $SU(2)$-invariant states
is studied and the paper consist of the following sections: In section II we review the definition of LQU for bipartite systems and in section III we show the definition of
$SU(2)$ invariant states. Finally, in section IV we have calculated the
LQU of a spin-1/2 particle and arbitrary spin-$j$ particle or of a spin-1 particle and a spin-$j$ particle with $SU(2)$ symmetry.
The paper is ended with a brief conclusion.

\section{Local quantum uncertainty}
Classically, it is possible to measure any two observables
with arbitrary accuracy. However, such measurement is
not always possible in quantum systems. Uncertainty
relation gives the statistical nature of errors in these kinds
of measurement. Measurement of single observable can
also help to detect uncertainty of a quantum observable.
For a quantum state $\rho$, an observable is called quantum
certain, if the error in measurement of the observable is
due to only the ignorance about the classical mixing in
$\rho$. A good quantifier of this uncertainty to an observable
is the skew information, defined by Wigner and Yanase \cite{Wigner} as
\begin{eqnarray}
I(\rho,K):=\frac{1}{2}Tr{[\sqrt{\rho},K^{A}]^{2}}
\end{eqnarray}
For a bipartite quantum state
$\rho_{AB}$, Girolami et.al. \cite{Girolami} introduced the concept of local
quantum uncertainty (LQU) defined as
\begin{eqnarray}
\mathcal{U}_{A}:=min_{K^{A}} I(\rho_{AB},K^{A})
\end{eqnarray}
The minimization is performed over all local maximally
informative observable (or non-degenerate) $K^{A}=K_{A}\otimes I$.
This quantity quantifies the minimum amount of uncertainty in a quantum state.

The closed form of the LQU for $2\otimes d$ quantum systems as introduced by Girolami \cite{Girolami} is
\begin{eqnarray}
\mathcal{U}_{A}:=1-\lambda_{max}(W)
\end{eqnarray}
In which $\lambda_{max}$ is the maximum eigenvalue of the $3 \times 3$ matrix W with the elements $W_{ij}=Tr{\sqrt{\rho} (\sigma_{i} \otimes I) \sqrt{\rho} (\sigma_{j} \otimes I)}$
and $\sigma_{i}$, $i = 1, 2, 3$ is the Pauli matrices.

\section{SU(2)-invariant states}
Considering two particles, one with spin $j_{1}$ and the other $j_{2}$, and the states that
are symmetric under global rotations. The symmetry group is $R=\{D^{j_{1}}(R)\otimes D^{j_{2}}(R)\}$, where $D^{j}(R)=exp(-i\theta \vec{J}.\vec{n})$ and
J is the angular-momentum vector. The twirling operator can be shown
\begin{eqnarray}
P_{R}(\rho)=\int d\mu(R)D^{j_{1}}(R)\otimes D^{j_{2}}(R)\rho D^{j_{1}}(R)^{\dagger}\otimes D^{j_{2}}(R)^{\dagger}
\end{eqnarray}
where, $\mu(R)$ is the group-invariant measure for the rotation group. The states that are invariant under the twirling operation are those that are convex
combinations of the states associated with the projectors onto the irreducible subspaces of
total angular momentum. These projectors are
\begin{eqnarray}
\Pi_{J}=\sum_{m=-J}^{m=J}\left|J,m\right\rangle \left\langle J,m\right|
\end{eqnarray}
Thus the $R$-invariant states are
\begin{eqnarray}
\rho(P)=\sum_{J=|j_{1}-j_{2}|}^{J=j_{1}+j_{2}} \frac{p_{J}}{2J+1}\Pi_{J}
\end{eqnarray}
where $p_{J}=Tr(\rho(P) \Pi_{J})\geq 0 $ and $\sum p_{J}=1 $ \cite{Kiran}.

\section{Main Discussion}
\subsection{spin-$j$ and spin-1/2}
An SU(2)-invariant state $\rho$ of a bipartite system that composed of a spin-$j$ and a spin-1/2 subsystems
are parameterized by a single parameter which will be denoted by $P$ throughout this paper and
can be written in total spin basis as \cite{Fanchini}

\begin{eqnarray}
\rho^{AB}=\frac{p}{2j} \sum_{m=-j+1/2}^{j-1/2} \left|j-1/2,m\right\rangle \left\langle j-1/2,m\right|+\frac{1-p}{2(j+1)} \sum_{m=-j-1/2}^{j+1/2} \left|j+1/2,m\right\rangle \left\langle j+1/2,m\right|
\end{eqnarray}
The eigenstates of total angular momentum are well known, given by the Clebsch-Gordon coefficients and can be written as
\begin{eqnarray}
\left|j\pm1/2,m\right\rangle=a_{\pm}\left|j,m-1/2\right\rangle \otimes \left|1/2,1/2\right\rangle+
b_{\pm}\left|j,m+1/2\right\rangle \otimes \left|1/2,-1/2\right\rangle
\end{eqnarray}
where $a_{\pm}=\pm \sqrt{\frac{j+1/2\pm m}{2j+1}} $ and $b_{\pm}=\sqrt{\frac{j+1/2\mp m}{2j+1}} $. By substitution the Clebsch-Gordon coefficients in eq.4,
we can write the density matrix in product basis as

\begin{eqnarray}
\begin{array}{c}
 \rho^{AB}=\sum_{m=-j}^{j} \left|m\right\rangle \left\langle m\right|\otimes (u \left|1/2\right\rangle \left\langle 1/2\right|+
 v \left|-1/2\right\rangle \left\langle -1/2\right|)+ \\\\
 \sum_{m=-j}^{j} w (\left|m\right\rangle \left\langle m+1\right|\otimes \left|1/2\right\rangle \left\langle -1/2\right|+\left|m+1\right\rangle \left\langle m\right|\otimes \left|-1/2\right\rangle \left\langle 1/2\right|)
 \end{array}
\end{eqnarray}
where
\begin{eqnarray}
\begin{array}{c}
 u=\frac{P}{2j}(\frac{j-m}{2j+1})+\frac{1-P}{2(j+1)}(\frac{j+m+1}{2j+1}) \\\\
 v=\frac{P}{2j}(\frac{j+m}{2j+1})+\frac{1-P}{2(j+1)}(\frac{j-m+1}{2j+1}) \\\\
 w=-\frac{\sqrt{(j-m)(j+m+1)}}{2j+1}(\frac{P}{2j}-\frac{1-P}{2(j+1)})
\end{array}
\end{eqnarray}

It is straightforward to calculate the $\sqrt{\rho^{AB}}$ and to evaluate the LQU of $\rho^{AB}$, let $K^{B}=(I_{A}\otimes K_{B})$
denote a local observable, with $K_{B}=\vec{n}.\vec{\sigma}$ and $(|\vec{n}|=1)$, where $\sigma_{i}$, $i$=(1, 2, 3) represent the Pauli matrices, which are the generators
of $SU(2)$.  Since making the measurements on spin-$j$ subsystem
will make the minimization procedure even harder, thus local observables are made on the spin-1/2 subsystem or $B$ subsystems.
After calculations,
we find that LQU do not depend on the measurement parameters, therefore, our calculation do not require any minimization over the projective measurements.

Then for integer and half-integer $j$, we have

\begin{eqnarray}
\mathcal{U}_{A}=\frac{8j(j+1)(\sqrt{\frac{P}{2j}}-\sqrt{\frac{1-P}{2(j+1)}})^{2}}{3(2j+1)}
\end{eqnarray}
Fig.1 and Fig.2 shows the LQU of spin-$j$ and spin-1/2 system for some half-integer and integer $j$, respectively.
It is important to note that as $P$=$j$/(2$j$+1) the LQU is exactly zero and
as we can see, for high dimensional systems (large j), the LQU becomes symmetric around the point $P$=1/2 where LQU vanishes.
Also, as we can see in Fig.1, when $j$=1/2, the state becomes the $2\otimes 2$ Werner state, and the LQU is equal to 1 at $P$=1.

\subsection{spin-$j$ and spin-1}
An SU(2)-invariant state $\rho$ of a bipartite system that is composed of a spin-$j$ and a spin-1 subsystems
are parameterized by parameters denoted by $P$ and $Q$ is \cite{Fanchini}

\begin{eqnarray}
\begin{array}{c}
\rho^{AB}=\frac{P}{2j-1} \sum_{m=-j+1}^{j-1} \left|j-1,m\right\rangle \left\langle j-1,m\right|+\frac{Q}{2j+1}
\sum_{m=-j}^{j} \left|j,m\right\rangle \left\langle j,m\right|+ \\\\
\frac{1-P-Q}{2j+3} \sum_{m=-j-1}^{j+1} \left|j+1,m\right\rangle \left\langle j+1,m\right|
\end{array}
\end{eqnarray}

The eigenstates of total angular momentum are well known, given by the Clebsch-Gordon coefficients and can be written as
\begin{eqnarray}
\begin{array}{c}
\left|j+1,m\right\rangle=a_{1}\left|m-1\right\rangle \otimes \left|1\right\rangle+a_{2}\left|m\right\rangle \otimes \left|0\right\rangle+
a_{3}\left|m+1\right\rangle \otimes \left|-1\right\rangle\\\\
\left|j,m\right\rangle=b_{1}\left|m-1\right\rangle \otimes \left|1\right\rangle+b_{2}\left|m\right\rangle \otimes \left|0\right\rangle+
b_{3}\left|m+1\right\rangle \otimes \left|-1\right\rangle\\\\
\left|j-1,m\right\rangle=c_{1}\left|m-1\right\rangle \otimes \left|1\right\rangle+c_{2}\left|m\right\rangle \otimes \left|0\right\rangle+
c_{3}\left|m+1\right\rangle \otimes \left|-1\right\rangle
\end{array}
\end{eqnarray}
where
\begin{eqnarray}
\begin{array}{c}
a_{1}=\sqrt{\frac{(j+m)(j+m+1)}{(2j+1)(2j+2)}}, a_{2}=\sqrt{\frac{(j-m+1)(j+m+1)}{(2j+1)(j+1)}}, a_{3}=\sqrt{\frac{(j-m)(j-m+1)}{(2j+1)(2j+2)}} \\\\
b_{1}=-\sqrt{\frac{(j+m)(j-m+1)}{2j(j+1)}}, b_{2}=\frac{m}{\sqrt{j(j+1)}}, b_{3}=\sqrt{\frac{(j-m)(j+m+1)}{2j(j+1)}} \\\\
c_{1}=\sqrt{\frac{(j-m)(j-m+1)}{2j(2j+1)}}, c_{2}=-\sqrt{\frac{(j-m)(j+m)}{j(2j+1)}}, c_{3}=\sqrt{\frac{(j+m)(j+m+1)}{2j(2j+1)}}.
\end{array}
\end{eqnarray}
By using the Clebsch-Gordon coefficients, density matrix in product basis can be written as
\begin{eqnarray}
\begin{array}{c}
 \rho^{AB}=\sum_{m=-j}^{j} \left|m\right\rangle \left\langle m\right|\otimes (v_{1}\left|1\right\rangle \left\langle 1\right|+v_{2}\left|0\right\rangle \left\langle 0\right|+v_{3}\left|-1\right\rangle \left\langle -1\right|)+ \\\\
 \sum_{m=-j}^{j} u_{1}(\left|m\right\rangle \left\langle m-1\right|\otimes \left|0\right\rangle \left\langle 1\right|+\left|m-1\right\rangle \left\langle m\right|\otimes \left|1\right\rangle \left\langle 0\right|) \\\\
 \sum_{m=-j}^{j} u_{2}(\left|m\right\rangle \left\langle m+1\right|\otimes \left|0\right\rangle \left\langle -1\right|+\left|m+1\right\rangle \left\langle m\right|\otimes \left|-1\right\rangle \left\langle 0\right|) \\\\
 \sum_{m=-j}^{j} u_{3}(\left|m-1\right\rangle \left\langle m+1\right|\otimes \left|1\right\rangle \left\langle -1\right|+\left|m+1\right\rangle \left\langle m-1\right|\otimes \left|-1\right\rangle \left\langle 1\right|)
 \end{array}
\end{eqnarray}
where
\begin{eqnarray}
\begin{array}{c}
v_{1}=\sqrt{\frac{P}{2j-1}}c_{1}^{2}+\sqrt{\frac{Q}{2j+1}}b_{1}^{2}+\sqrt{\frac{1-P-Q}{2j+3}}a_{1}^{2} \\\\
v_{2}=\sqrt{\frac{P}{2j-1}}c_{2}^{2}+\sqrt{\frac{Q}{2j+1}}b_{2}^{2}+\sqrt{\frac{1-P-Q}{2j+3}}a_{2}^{2} \\\\
v_{3}=\sqrt{\frac{P}{2j-1}}c_{3}^{2}+\sqrt{\frac{Q}{2j+1}}b_{3}^{2}+\sqrt{\frac{1-P-Q}{2j+3}}a_{3}^{2} \\\\
u_{1}=\sqrt{\frac{P}{2j-1}}c_{1}c_{2}+\sqrt{\frac{Q}{2j+1}}b_{1}b_{2}+\sqrt{\frac{1-P-Q}{2j+3}}a_{1}a_{2} \\\\
u_{2}=\sqrt{\frac{P}{2j-1}}c_{3}c_{2}+\sqrt{\frac{Q}{2j+1}}b_{3}b_{2}+\sqrt{\frac{1-P-Q}{2j+3}}a_{3}a_{2} \\\\
u_{3}=\sqrt{\frac{P}{2j-1}}c_{1}c_{3}+\sqrt{\frac{Q}{2j+1}}b_{1}b_{3}+\sqrt{\frac{1-P-Q}{2j+3}}a_{1}a_{3}
\end{array}
\end{eqnarray}
Now, let $K^{B}=(I_{A}\otimes K_{B})$ denote a local observable, with $K_{B}=\vec{n}.\vec{\lambda}$ and $(|\vec{n}|=1)$, where $\lambda_{i}$, with $i$=(1, 2, 3, 4, 5, 6, 7, 8) represent the generators of $SU(3)$. Since the LQU definition requires a minimization over the local observable, then we need to optimize the LQU value over the local observable.
The procedure of our approach is straightforward, we used the method of Lagrange multipliers finding the local maxima and minima of a function subject to equality constraint. Firstly,
we find the optimum value of the LQU is achieved when $(n1,n2,n3,n4,n5,n6,n7,n8)=(0,0,\pm 1/2,0,,0,0,0,\pm \sqrt{3}/2)$. This shows that if the local measurements are performed by them, then the related LQU are obtained as: $\emph{u}_{A}=\sum_{m=-j}^{j}(u_{1}^2+u_{2}^2+4u_{3}^2)$.
The other optimum value of LQU is achieved when
$(n1,n2,n3,n4,n5,n6,n7,n8)=(0,0,\mp \sqrt{3}/2,0,0,0,0,\pm 1/2)$, yielding the LQU as: $\emph{u}_{A}=3\sum_{m=-j}^{j}(u_{1}^2+u_{2}^2)$.
Now, we have;
\begin{eqnarray}
\mathcal{U}_{A}=min(\sum_{m=-j}^{m=j}(u_{1}^2+u_{2}^2+4u_{3}^2), 3\sum_{m=-j}^{m=j}(u_{1}^2+u_{2}^2))
\end{eqnarray}
In Fig.3, Fig.4 and Fig.5 we show the LQU of spin-$j$ and spin-1 system that is a function of $P$ and $Q$ for $j=1,5/2$ and $j=10$, respectively.

\section{Conclusions}
Recently, some measures such as the entanglement of formation (EoF), I-concurrence, I-tangle, convex-roof-extended negativity and Quantum discord 
of the $SU(2)$-invariant states of a spin-$j$ and spin-1/2 particles have been analytically calculated.
In this paper we have studied a wider range of $SU(2)$ invariant states that consisting of
a spin-$j$ and a spin-1/2 subsystems and also a spin-$j$ and a spin-1 subsystems by the LQU and
we have derived analytical expression for them.
However, we suggest that the LQU for other states (or states being invariant under other
transformation groups) can be studied in the future investigation.

\newpage
Fig. 1. LQU of the bipartite state composed of a spin-$j$ and a spin-1/2 vs. $P$  for $j$ = 1/2 (d = 2), $j$ = 5/2 (d = 6) and $j$ = 101/2 (d = 102).
\begin{figure}
\centering
\includegraphics[width=445 pt]{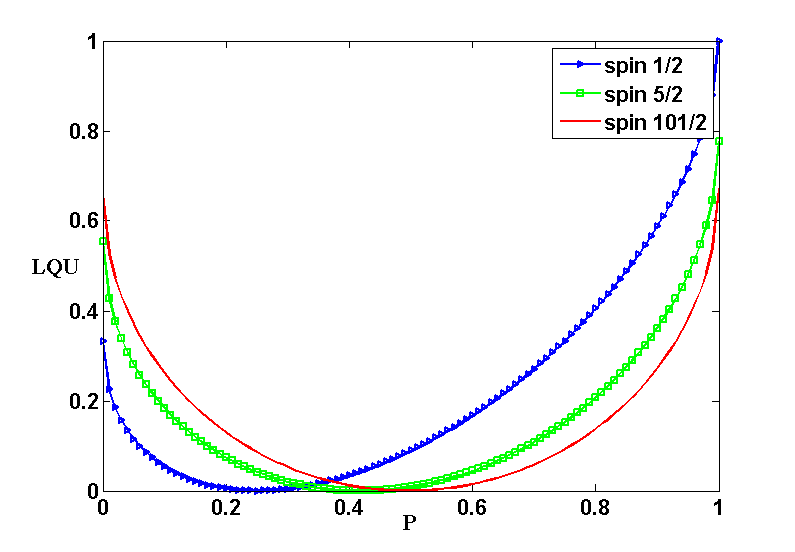}
\caption{} \label{Fig}
\end{figure}

\newpage
Fig. 2. LQU of the bipartite state composed of a spin-$j$ and a spin-1/2 vs. $P$  for $j$ = 1 (d = 3), $j$ = 5 (d = 11) and $j$ = 100 (d = 201).
\begin{figure}
\centering
\includegraphics[width=445 pt]{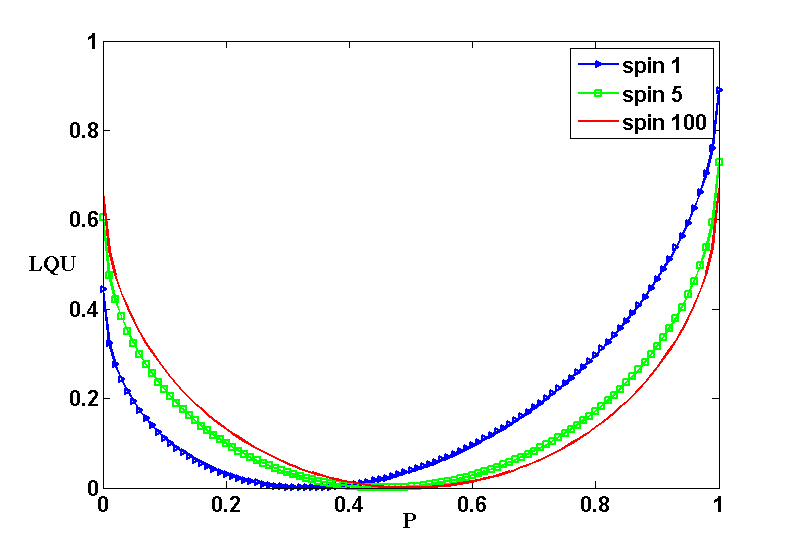}
\caption{} \label{Fig}
\end{figure}

\newpage
Fig. 3. LQU of the bipartite state composed of a spin-$j$ and a spin-1 as a function of $P$ and $Q$ for $j$ = 1 (d = 3).
\begin{figure}
\centering
\includegraphics[width=445 pt]{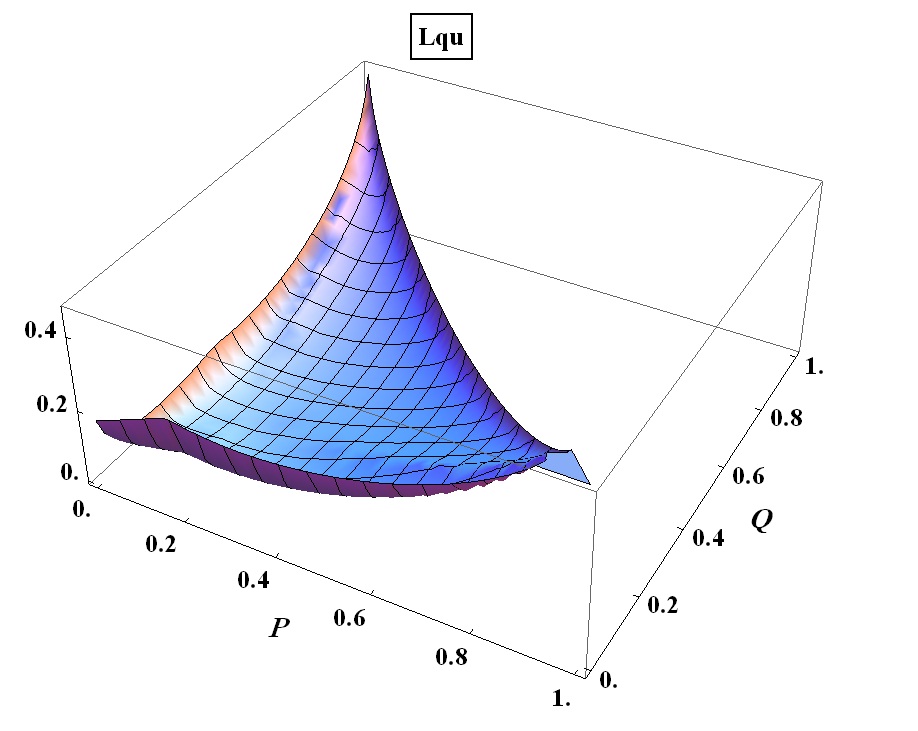}
\caption{} \label{Fig}
\end{figure}

\newpage
Fig. 4. LQU of the bipartite state composed of a spin-$j$ and a spin-1 as a function of $P$ and $Q$ for $j$ = 5/2 (d = 6).
\begin{figure}
\centering
\includegraphics[width=445 pt]{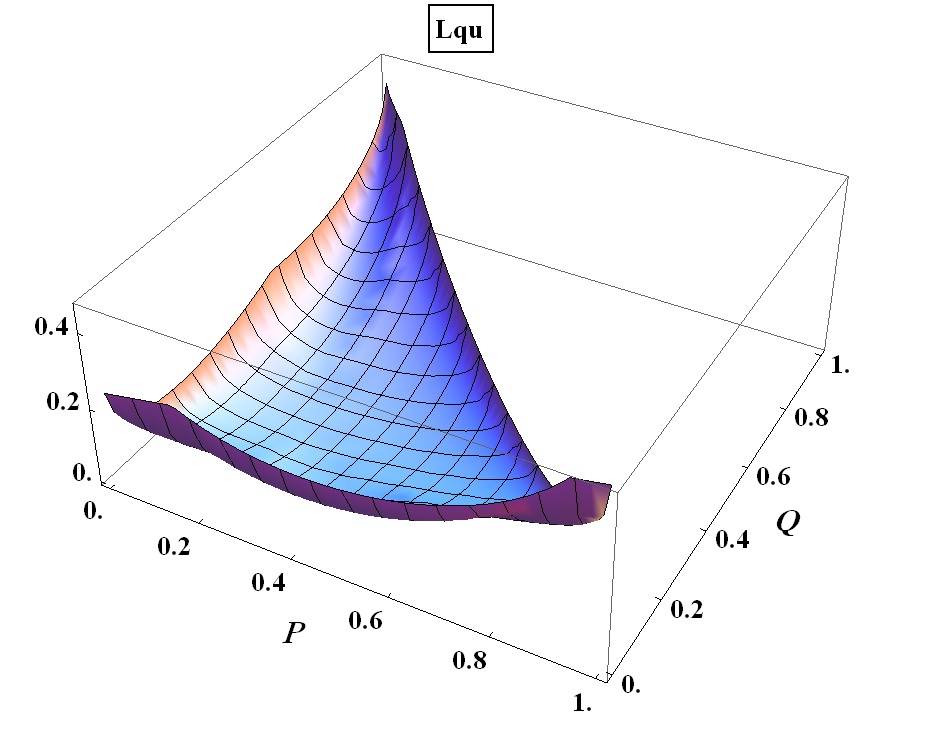}
\caption{} \label{Fig}
\end{figure}

\newpage
Fig. 5. LQU of the bipartite state composed of a spin-$j$ and a spin-1 as a function of $P$ and $Q$ for $j$ = 10 (d = 21).
\begin{figure}
\centering
\includegraphics[width=445 pt]{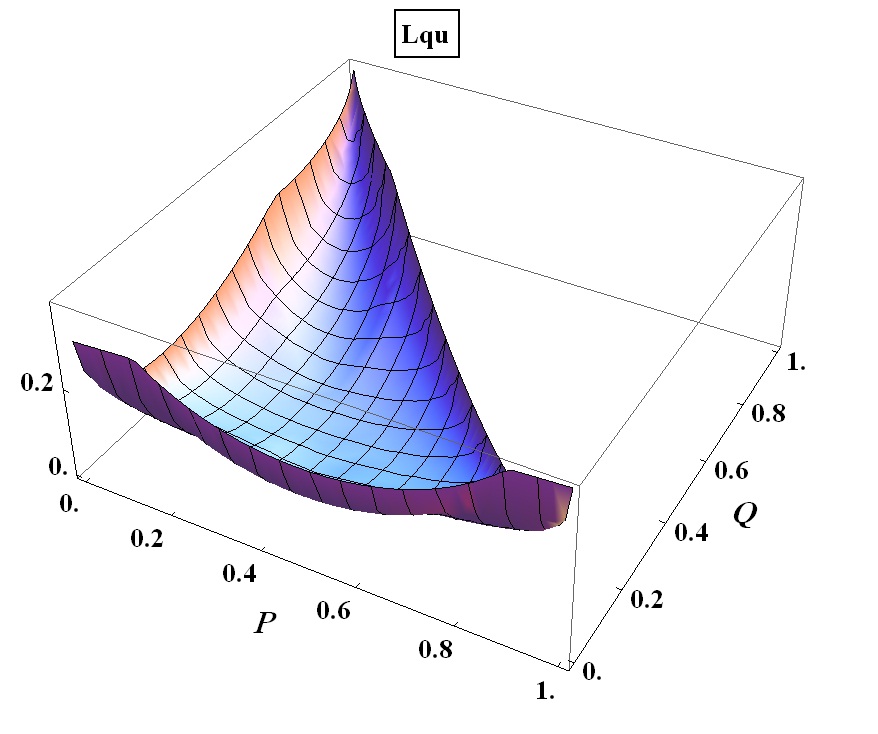}
\caption{} \label{Fig}
\end{figure}


\newpage
\begin{thebibliography}{99}

\bibitem{Bennett}
C. H. Bennett, D. P. DiVincenzo, C. A. Fuchs, T. Mor, E. Rains, P . W. Shor, J. A. Smolin and W.
K. Wootters, Phys. Rev. A 59, 1070 ( 1999).

\bibitem{Henderson}
L. Henderson and V. Vedral, J. Phys. A 34, 6899 (2001).

\bibitem{Modi}
K. Modi, A. Brodutch, H. Cable, T. Paterek, and V. Vedral, Rev. Mod. Phys. 84, 1655 (2012).

\bibitem{Ollivier}
H. Ollivier and W. H. Zurek, Phys. Rev. Lett. 88, 017901 (2001).

\bibitem{Dakic}
Dakic, Vedral, and Brukner, Phys. Rev. Lett. 105, 190502
(2010).

\bibitem{Bellomo}
B. Bellomo, R. Lo Franco, and G. Compagno, Phys. Rev. A 86,
012312 (2012).

\bibitem{Rajagopal}
A. K. Rajagopal and R.W. Rendell, Phys. Rev. A 66, 022104 (2002).

\bibitem{Luo1}
S. Luo, Phys. Rev. A 77, 022301 (2008).

\bibitem{Girolami}
D. Girolami, T. Tufarelli and G. Adesso, Phys. Rev. Lett. 110, 240402 (2013).

\bibitem{Vinjanampathy}
S. Vinjanampathy, A. R. P. Rau, J. Phys. A 45, 095303 (2012).

\bibitem{Shuhao}
Shuhao Wang, Hui Li, Xian Lu and Gui-Lu Long arXiv:1307.0576 [quant-ph].

\bibitem{Schliemann1}
J. Schliemann, Phys. Rev. A 68, 012309 (2003).

\bibitem{Schliemann2}
J. Schliemann, Phys. Rev. A 72, 012307 (2005).

\bibitem{Wang}
Z. Wang and Z. X. Wang, Phys. Lett. A 372, 7033 (2008).


\bibitem{Manne}
K. K. Manne, C. M. Caves, Quantum. Inf. Comp. 8, 0295 (2008).

\bibitem{Vollbrecht}
K. G. H. Vollbrecht and R. F. Werner, Phys. Rev. A 64, 062307 (2001).

\bibitem{Cakmak}
B. Cakmak and Z. Gedik, J. Phys. A: Math. Theor. 46 (2013) 465302.

\bibitem{Teng}
Yao-Kun Wang, Teng Ma, Shao-Ming Fei, and Zhi-Xi Wang, arXiv:1307.3468 [quant-ph].

\bibitem{Wigner}
E. P. Wigner, M. M. Yanase, Proc. Natl. Acad. Sci. U.S.A. 49, 910-918 (1963).

\bibitem{Kiran}
Kiran K. Manne and Carlton M. Caves, arXiv:0506151 [quant-ph].

\bibitem{Fanchini}
F. F. Fanchini, L. K. Castelano, M. F. Cornelio, and M. C. de Oliviera, New J. Phys. 14, 013027 (2012).


\end{thebibliography}
\end{document}